\documentclass[a4paper,20pt]{article}
    \usepackage[T1]{fontenc}
    \usepackage{epsfig}
    \usepackage{graphicx}
    \usepackage{amsmath}
    \usepackage{amsfonts}
    \renewcommand{\abstract}{}
\begin{document}
\makeatletter
\renewcommand{\@oddhead}{\textit{YSC'14 Proceedings of Contributed Papers} \hfil \textit{M. Dyrka, B. Wszo\l ek}}
\renewcommand{\@evenfoot}{\hfil \thepage \hfil}
\renewcommand{\@oddfoot}{\hfil \thepage \hfil}
\fontsize{11}{11} \selectfont

\title{Interstellar $C_2$ Molecule Detected in UV Spectra of Reddened Stars}
\author{\textsl{M. Dyrka, B. Wszo\l ek}}
\date{}
\maketitle
\begin{center} {Jan D\l ugosz Academy, Institute of Physics, al. Armii
Krajowej 13/15, 42-200 Cz\k{e}stochowa, Poland \\ bogdan@ajd.czest.pl}
\end{center}

\begin{abstract}
$C_2$ molecule is sometimes considered as a crucial component of carriers of some diffuse interstellar bands. Using UV data achieved by spectrometer  STIS fed with HST we detected interstellar $C_2$ lines for few reddened
target stars. We tried to verify the idea that intensity of $C_2$ lines  around 2313 \AA \, is correlated with some diffuse interstellar bands.
\end{abstract}

\section*{Introduction}
\indent \indent The identification of the carriers of diffuse interstellar bands (DIBs) has been a long standing problem in astrophysics \cite{herbig95}. The spectra of DIBs have the characteristics of both solid state carriers and free gas phase molecules. Many candidates have been proposed for the carriers of DIBs. Among them are carbon chain molecules. After hydrogen, carbon is the most important constituent of the interstellar  medium. Spectroscopic observations of early spectral type stars, embedded in, or behind diffuse interstellar clouds, reveal absorption lines of diatomic molecules such as CN, CH and $C_2$. It has been argued  that electronic transitions of carbon chains may be among the carriers of unidentified absorption features in diffuse interstellar clouds (see e.g. \cite{thaddeus}). Particularly longer chains are of interest as these are expected to be photo stable.

In 1977 Douglas \cite{douglas} suggested that polycarbon chains such as $C_5$, $C_7$ or $C_9$
might survive under interstellar conditions. He remarked that the means  by which it might happen is the internal conversion. That process enables the radiationless transition to take place from the excited
level to vibrational levels of the same or another electronic state, following which the absorber returns to the ground state by a series  of infrared transitions. Douglas also suggested that such carbon chains should produce strong diffuse absorption bands in the 4000-5000 \AA \,region. These suggestions were firstly pursued in
the laboratory by Kratschmer et al. \cite{krautschmer} and subsequently by Kurtz and Huffman \cite{kurtz}. Carbon molecules obtained by vaporization of graphite ($C_n$ , where n=4-9) and then deposited into a cold argon
matrix produced a series of absorption bands centred at 2471 \AA (the strongest), 3079, 3490, 4470 \AA (weaker) and 4930, 5860 \AA (very weak). Since the matrix shift estimated by Kratschmer et al. is about 50 \AA, the band 4470 matches DIB 4428 very well.

In 1998 Tulej et.al. \cite{tulej} measured the gas-phase electronic spectra of several carbon chain anions. They obtained spectra of $C_6^-$, $C_7^-$, $C_8^-$ and $C_9^-$ containing many narrow bands which appear to match some DIBs. Most intriguing coincidence was found for the bands of $C_7^-$ anion and several DIBs where the DIBs have similar widths and relative intensities. Mitchell and Huntress \cite{mitchell} went a bit further and proposed hydrocarbon chain molecules of up to $C_{12}$ $H_m$ (where m was unspecified)  as the species that might be responsible for at least some of the DIBs.  In 1994 Freivogel et al. \cite{freivogel} obtained the spectra of $C_n$ $H_m^-$ anions which show many correspondences with some DIBs. Particularly striking are
coincidences in the region 7000-8000 \AA \, where nine bands match relatively  well DIBs listed by Herbig and Leka \cite{herbig91}.

All results above cited were achieved by physical chemists. They did not meet full acceptance of astronomers who analyse astronomical spectra. Unfortunately, all the suggested matchings of bands are not  as tight as needed. But astronomers observe short carbon chains in interstellar clouds. That fact may support the idea that long carbon chains are also present there and they play a role of DIB carriers. Phillips system (around 10145 \AA) of the interstellar diatomic carbon  has been discovered by Souza and Lutz \cite{souza} toward IV Cyg 12.

The F-X (1342 \AA) band of $C_2$ was detected by Lien \cite{lien} toward X Per. Using echelle spectra acquired by Goddard High Resolution Spectrograph (aboard the Hubble Space Telescope) D-X (2313 \AA) and F-X (1342 \AA) bands of $C_2$ were detected toward X Per by Kaczmarek \cite{kaczmerek}. Ka\'zmierczak et al. \cite{kazmierczak}
found interstellar lines of $C_2$ (Phyllips bands) towards 6 early-type stars. In this contribution we will present our results of searching for interstellar $C_2$ molecule in the UV domain.

\section*{Observational data}

\indent \indent We have used spectroscopic UV data from Space
Telescope Imaging Spectrograph (STIS). We took STIS spectra of
early-type reddened stars via internet visiting homepage:
http://archive.stsci.edu/. STIS archive does not contain spectra of
all target stars used for DIB exploration. We have found UV spectra
with appropriate wavelength ranges (where $C_2$ manifests its
presence) only for 12 stars. These stars are (HD): 22591, 23180,
24398, 24534, 27778, 34078, 147933, 192639, 198478, 206267, 207198
and 210839. Reddenings, E(B-V), of these stars are between 0.23 and
1.69.  Echelle spectra are accessible as binary FITS files. Mostly,
observations are multiple. We used high resolution spectra with
R=110000.

\section*{Data analysis}
\indent \indent There are three electron transition systems of the $C_2$ molecule which are accessible with STIS spectrograph:
\begin{itemize}
\item D-X system with its 0-0 band near 2313 \AA,
\item F-X system with 0-0 band near 1342 \AA,
\item F-X system with 1-0 band near 1314 \AA.
\end{itemize}
We focused our attention on all of these line systems. Each system contains a set of numerous and well separated rotational lines. Each star from our sample was observed for several times. To get S/N ratio as good as possible, we averaged observations. Figures 1-3 show the results of our analysis. Considered $C_2$ bands are clearly visible only for few target stars. For the other few ones we could find only very weak trace of structure. D-X system is the most prominent here.

Unfortunately, STIS spectra were accessible for 4 our stars only. In much weaker system around 1342 \AA \,we found expected structure, or their trace, for 7 stars from our sample. These stars are (HD): 23180, 24534, 27778, 147933, 206267, 207198 and 210839. Finally, we tried to look for correlation between intensities of $C_2$ lines and DIBs (5780, 5797). Data for DIBs we took from literature \cite{wszolek}. The amount of diatomic carbon does not seem to correlate with DIB carriers' abundances nor with reddening. For 5780 DIB one can notice only that its carrier tends to avoid regions, where the $C_2$ molecule is very abundant.

\section*{Concluding remarks}

\indent \indent One of the results of our analysis is one more
confirmation that  $C_2$ molecules are present in some interstellar
clouds producing DIBs. Furthermore, we showed that not only 2313 \AA
\,band of $C_2$, but also 1342 and 1314 \AA \,bands, may be
detectable in STIS data.

To solve the problem whether $C_2$ may be crucial molecule as far as DIBs' carriers are concerned, one needs much more observations of $C_2$ lines. As we know today, it is hard to argue that simple carbon chains or their
ions can play a role of some DIBs' cariers. However, $C_2$ molecule is often considered by researchers as a crucial molecule for rising carriers of some DIBs. Our results do not give us right to claim that column density of interstellar $C_2$ molecule is well correlated with strengths of strong DIBs. However, the tendency noticed for 5780 DIB carrier, that it seems to avoid regions with big concentrations of $C_2$, is very interesting to be verified.

\newpage

\textbf{Figure 1.} Averaged STIS spectra including F-X  1-0 band
system of $C_2$  molecule (1314 \AA).\vspace{10ex}

\textbf{Figure 2.} Averaged STIS spectra including F-X  0-0 band
system of $C_2$  molecule (1342 \AA).\vspace{10ex}

\textbf{Figure 3.} Averaged STIS spectra including D-X  0-0 band
system of $C_2$  molecule (2313 \AA).\vspace{10ex}

Figures are available on YSC home page
(http://ysc.kiev.ua/abs/proc14$\_$6.pdf).

\end{document}